\documentclass[a4paper,
               keeplastbox,   % flushend option: not to un-indent last line in References
               ]{jacow}
\usepackage{pdfpages,multirow,ragged2e} %
\makeatletter%
	\ifboolexpr{bool{xetex}}
	 {\renewcommand{\Gin@extensions}{.pdf,%
	                    .png,.jpg,.bmp,.pict,.tif,.psd,.mac,.sga,.tga,.gif,%
	                    .eps,.ps,%
	                    }}{}
\makeatother

\ifboolexpr{bool{xetex} or bool{luatex}} % test for XeTeX/LuaTeX
 {}                                      % input encoding is utf8 by default
 {\usepackage[utf8]{inputenc}}           % switch to utf8

\usepackage[USenglish]{babel}

\ifboolexpr{bool{jacowbiblatex}}%
 {%
  \addbibresource{jacow-test.bib}
  \addbibresource{biblatex-examples.bib}
 }{}
\begin{document}
\title{LLRF upgrade status at the KEK Photon Factory 2.5 GeV ring\thanks{Work supported by  JSPS KAKENHI Grant Number 20H04459.}}

\author{D. Naito\thanks{daichi.naito@kek.jp}, N. Yamamoto, T. Takahashi, A. Motomura, S. Sakanaka\\
            Accelerator Laboratory, High Energy Accelerator Research Organization(KEK), Tsukuba, Japan}
	
\maketitle

\begin{abstract}
In 2023, we are replacing the LLRF system for the KEK-PF 2.5 GeV ring. 
The new system is composed of digital boards such as eRTM, AMC, and $\mu$RTM, based on the MTCA.4 standard. 
In our system, we adopted the non-IQ direct sampling method for RF detection. 
We set the sampling frequency at 8/13 (307.75 MHz) of the RF frequency, where the denominator (13) is the divisor of the harmonic number (312) of the storage ring. 
This allows us to detect the transient variation of the cavity voltage that is synchronized with the beam revolution. 
We also plan to compensate for the voltage variation by implementing a feedforward technique.
These functions will be useful in a double RF system for KEK future synchrotron light source. 
Production and installation of the new system were complete and the new system is under commissioning.
In this presentation, we introduce our new system and report the upgrade status.
\end{abstract}

\section{Introduction}
At the KEK Photon Factory 2.5 GeV ring (PF ring), we are developing a new LLRF system to replace the current system~\cite{PFLLRF} by November 2023.
The current system is composed of analog modules, while the new system is composed of digital boards based on the MTCA.4 standard~\cite{MTCA}.
Table~\ref{RFP} shows principal parameters of the RF system at the PF ring.
The PF ring has two cavities in one section and a total of four cavities are installed in two sections.
Four klystrons are also installed to provide RF power to each cavity.
Since the original LLRF was designed 40 years ago, 
the system consists of old-fashioned analog circuit modules, some of which are becoming difficult to obtain.
To improve both the maintainability and the performance, we are developing the new LLRF system.
The new LLRF system also incorporates advanced technology such as bunch-phase detection and transient beam loading compensation for the future KEK light source~\cite{NY2018}.
In this paper, we introduce our new LLRF system and its features.
We also report the installation status and the performance tests of the new LLRF system.

\begin{table}[hbt]
\centering
\caption{The Principal Parameters of the RF System for the Photon Factory Storage Ring}
\begin{tabular}{lc}
\toprule
Parameter&Value \\
\midrule
Number of cavities & 4 \\
Radio frequency & 500.1 MHz\\
Harmonic number & 312\\
Cavity voltage per cavity & 0.425 MV\\
Beam current & 450 mA\\
Klystron power per cavity& 72 kW\\
\bottomrule
\end{tabular}
\label{RFP}
\end{table}

\section{New LLRF system}

\begin{figure}[hbt]
\centering
\includegraphics[width=80mm]{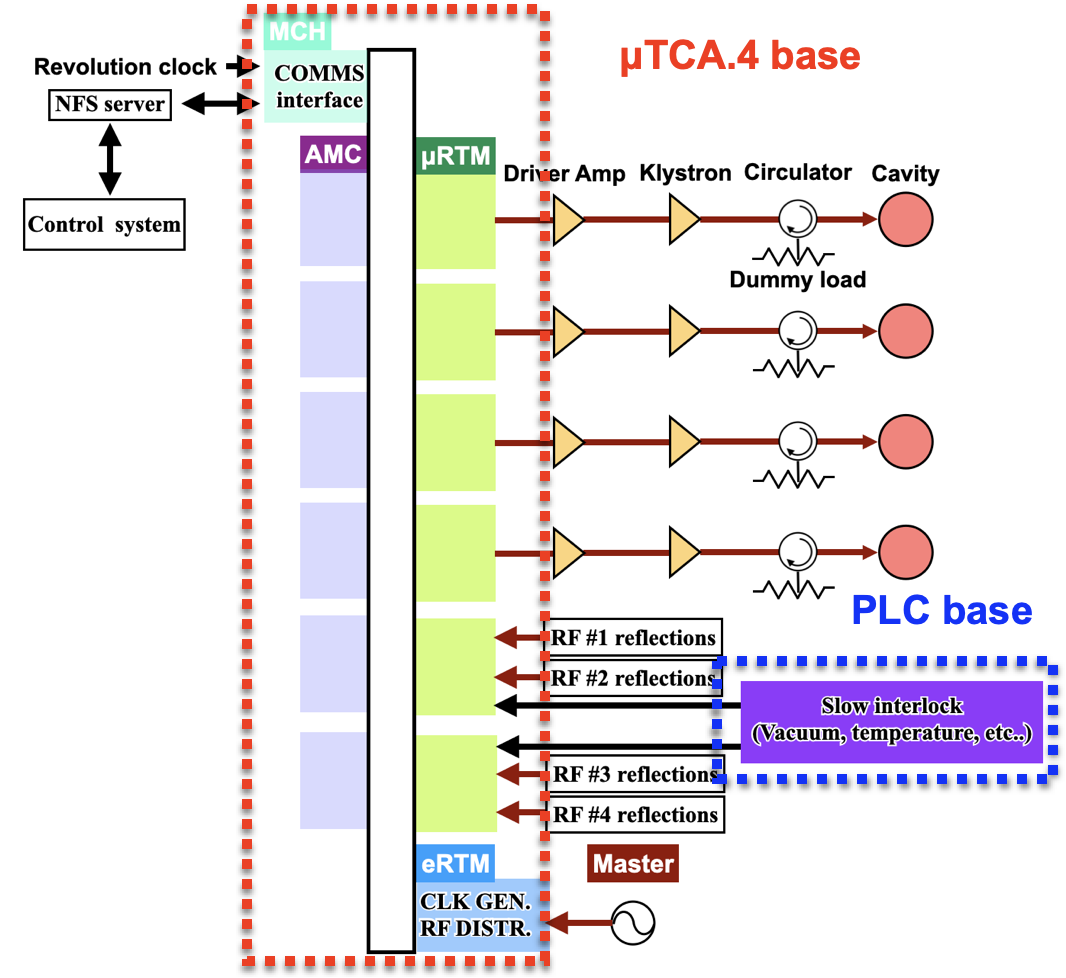}
\caption{Schematic of the new LLRF system.}
\label{DLLRF}
\end{figure}

\begin{figure}[hbt]
\centering
\includegraphics[width=60mm]{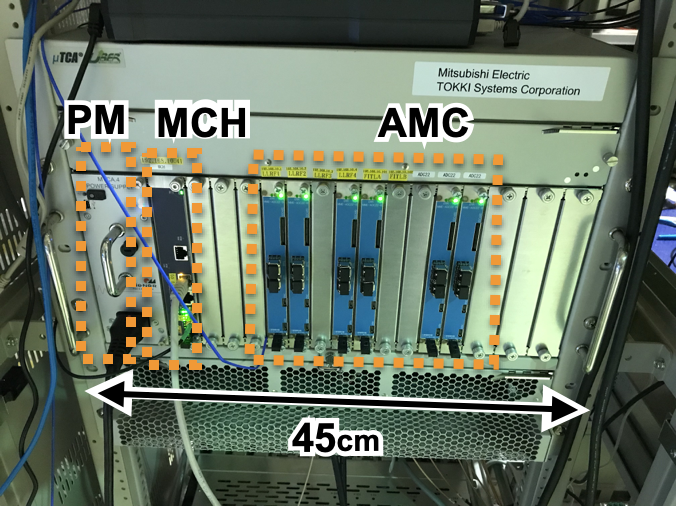}
\caption{Photo of the new LLRF system.}
\label{PDLLRF}
\end{figure}

Figures~\ref{DLLRF} and \ref{PDLLRF} show the schematic of the new LLRF system and its photo, respectively.
The new LLRF system is housed in a MTCA.4 shelf.
In the shelf, there are one Micro-TCA Carrier Hub (MCH), six pairs of Advanced Mezzanine Card (AMC) and Micro Rear Transition Module ($\mu$RTM),
and one Extended Rear Transition Module (eRTM).

The MCH mediates the EPICS communication between the AMCs and an operation server.  
Single pair of AMC and $\mu$RTM controls the single RF station and the four pairs are totally used for the RF control.
They are called LLRF boards.
The remaining pairs monitor the reflected signals from two RF stations per pair.
They are called the fast interlock (FITL) boards.
When the amplitude of the reflected signal exceeds to an interlock threshold, the FITL board sends the RF off signal to the LLRF boards via M-LVDS lines.
The eRTM generates a clock signal from the master RF signal and distributes it to each AMC via the RF backplane.
The eRTM also distributes the master RF signal to an IQ modulator located on the $\mu$RTM.
Table~\ref{DevPol} shows the development policy of each item used in the new LLRF.
Customizing the LLRF technologies developed for the J-PARC~\cite{AMC}, SPring-8~\cite{eRTM}, and SuperKEKB~\cite{KEKBLLRF}, 
we minimized the development costs, periods, and risks.
Our goal for the new LLRF design is to achieve the stability of cavity voltage within $\pm$0.1~\% in amplitude and $\pm$0.1~degrees in phase, respectively.

\begin{table}[hbt]
\centering
\caption{The Development Policy of Each Item Used in the New LLRF}
\begin{tabular}{lc}
\toprule
Item &Policy\\
\midrule
Shelf & Use commercial model\\
MCH & Use commercial model\\
AMC & Modify module developed at J-PARC\\
$\mu$RTM & Modify module developed at J-PARC\\
eRTM & Use module developed at SPring-8\\
Cavity stabilize& Same as  SPring-8\\
Cavity tuning & Same as SuperKEKB\\
\bottomrule
\end{tabular}
\label{DevPol}
\end{table}

\section{Features of the new LLRF system}
\begin{figure}[hbt]
\centering
\includegraphics[width=80mm]{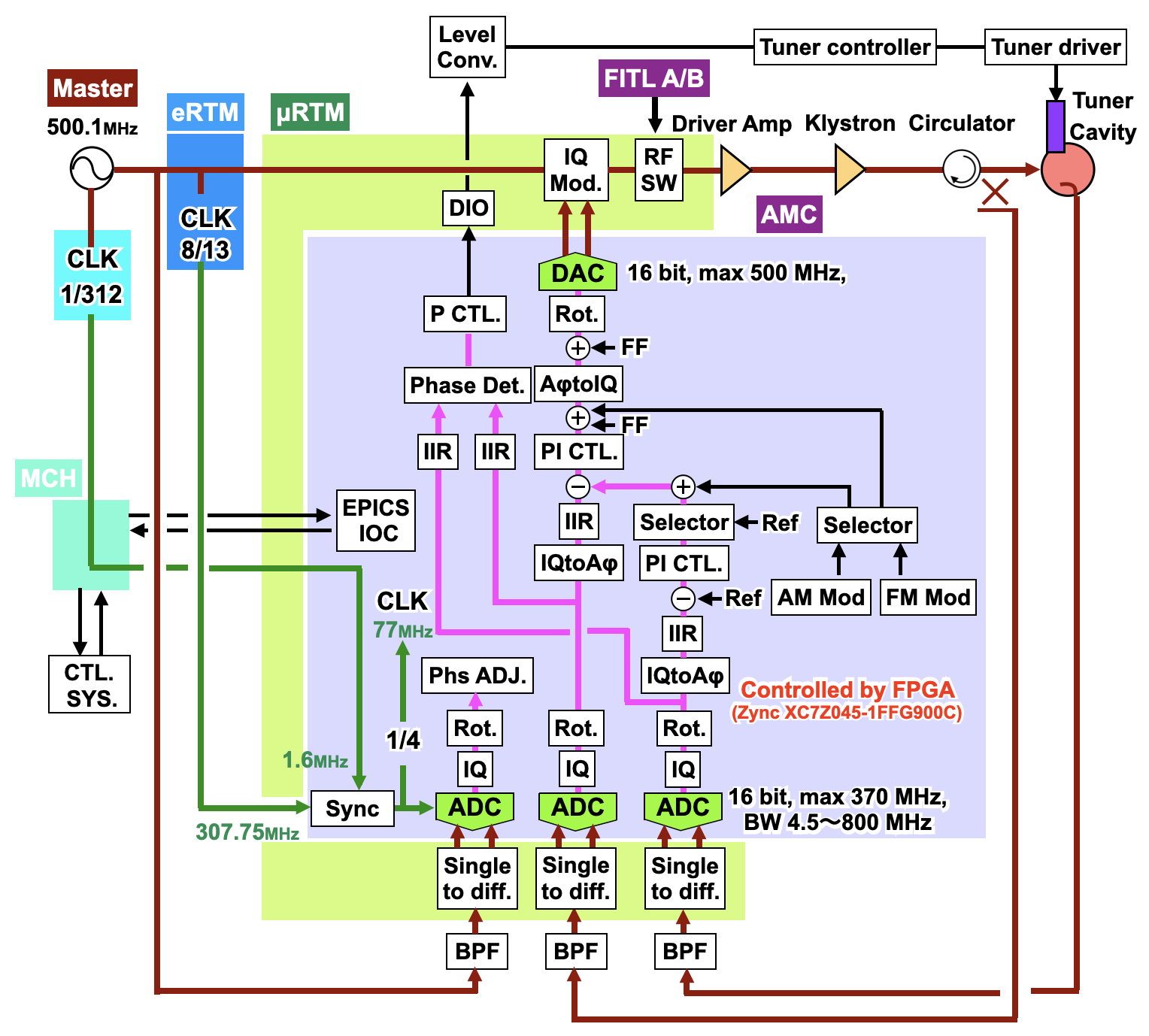}
\caption{RF control scheme of the new LLRF system.}
\label{RFCTL}
\end{figure}

Figure~\ref{RFCTL} shows the RF control scheme at the new LLRF system, which is executed on the FPGA and controlled by the EPICS communications.
The RF signal is detected by the ADC and output by the DAC via the IQ modulator.
The new scheme has mainly four features as described below:
\begin{enumerate}
\setlength{\itemsep}{0cm}
\item using non-IQ direct sampling for the RF detection
\item detecting signal variations with synchronization to the beam revolution frequency
\item stabilizing amplitude and phase with double feedback loops
\item modulating both the amplitude and phase of the RF output signal in arbitrary frequency
\end{enumerate}
In this section, we present the first, second, and third features.

\subsection{Non-IQ direct sampling and signal detection}
We adopted non-IQ direct sampling~\cite{IFS} for the RF detection of the ADC.
Using the voltage of the input RF signal ($V_i$) and the frequency of the RF ($f_{RF}$), 
the notation of our sampling are written as
\begin{equation}
I=\frac{1}{4}\sum^{7}_{i=0}V_i\cos\left(2\pi i\cdot\frac{5}{8}\right),
\end{equation}
\begin{equation}
Q=-\frac{1}{4}\sum^{7}_{i=0}V_i\sin\left(2\pi i\cdot\frac{5}{8}\right),
\end{equation}
\begin{equation}
f_s=f_{RF}\times\frac{8}{13}.
\end{equation}
We set the sampling frequency ($f_s$) at 8/13 (307.75 MHz) of the RF frequency, where the denominator (13) is the divisor of the harmonic number (312) of the storage ring. 
As a result, when the IQ signal is detected by exactly 24 times, the electron circulates the ring once.
This allows us to detect the transient variation of the cavity voltage that is synchronized with the beam revolution. 

\begin{figure}[hbt]
\centering
\includegraphics[width=70mm]{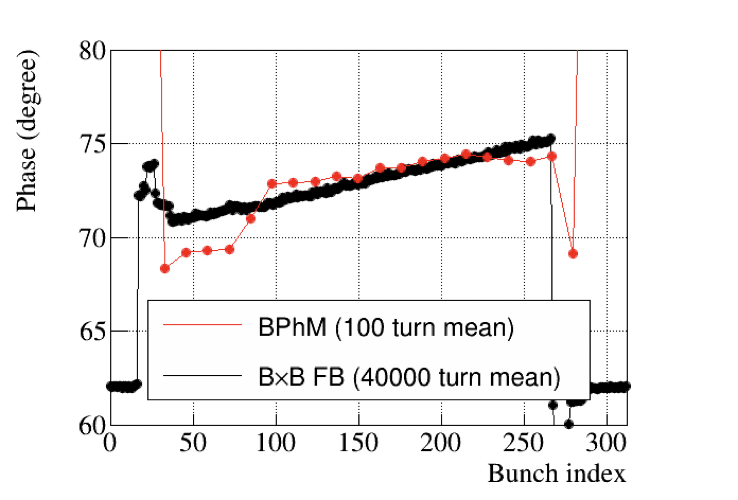}
\caption{The phase variation of the electron bunch measured by the digital LLRF and bunch by bunch feedback. 
The fill pattern consisted of 250 bunches and 62 empty RF buckets.}
\label{BPhase}
\end{figure}

Figure~\ref{BPhase} shows the transient variation of bunch phase which was measured with
the new LLRF and with a bunch by bunch feedback system (B$\times$B FB) installed in the PF~\cite{iGP}.
The black line shows the average phase of 40000 revolutions calculated by the B$\times$B FB.
The red line shows the average phase of 100 revolutions calculated by using a button-type pickup signal and the special firmware of the LLRF which calculates the average phase fast.
The horizontal and vertical axis show the bunch index and the phase at each bunch position, respectively.

Due to the transient RF voltage induced by a bunch gap, the synchronous phase of each bunch shifts with its position in the bunch train. 
In Fig.~\ref{BPhase},  the result from the new LLRF agreed reasonably with that from the B$\times$B FB.
The phase variation was successfully observed by the new LLRF.
In addition to the phase detection, we plan to compensate the phase variation by implementing a feedforward technique.
These functions will be useful in a double RF system for KEK future synchrotron light source~\cite{NY2018}. 

\subsection{Amplitude and phase control with double feedback loops}
As shown in Fig.~\ref{RFCTL}, the cavity voltage is stabilized using the cavity input and pickup signals.
The cavity input feedback corrects the ripples induced in the klystron power supply whose frequency is less than 10~kHz.
The cavity pickup feedback corrects the variation induced by tuning error and beam loading whose frequency is several Hz.
The output of the cavity pickup feedback is used as the reference of the cavity input feedback.
These feedbacks consist double feedback loops, which was proven at the SPring-8~\cite{eRTM}.

In both feedbacks, the RF signal is converted to the IQ signals as explained in the last section.
The IQ signals, then, are converted to the amplitude and phase by the CORDIC algorithm.
The amplitude and the phase are used as the inputs of the feedbacks.
Since the phase of the pickup signal rotates 65 degrees from the phase of the generator voltage with the beam current of 450~mA, we concerned that the feedback using IQ signals would be hard to manage in such situations.

%\begin{figure}[hbt]
%\centering
%\includegraphics[width=60mm]{Phaser.png}
%\caption{Phase diagram with the beam current of 450~mA.}
%\label{Phaser}
%\end{figure}

\section{Status of the installation and commissioning}
All of the hardware for the new LLRF system was produced in FY 2022 and installed in the PF by August 2023.
Tentative performance tests were almost completed by mid-October.
Currently, the RF system is under commissioning to establish the control parameters and procedures for the user operation.
Then, the commissioning with beam will begin in early November.
In this section, we report some results of the performance tests.

\subsection{Stability of the klystron output}
\begin{figure}[hbt]
\centering
\includegraphics[width=75mm]{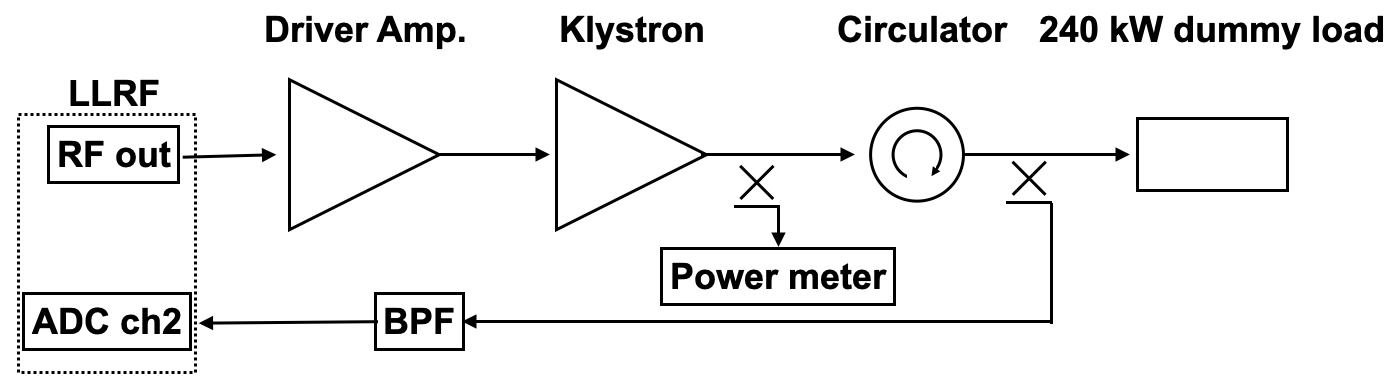}
\caption{Experimental setup of the klystron control test with the digital LLRF.}
\label{Klytest}
\end{figure}

We compared the power spectrum of the klystron output when the klystron was controlled by the current analog LLRF and the new digital LLRF.
Figure~\ref{Klytest} shows the experimental setup of the klystron control test with the digital LLRF.
Instead of the cavity, the 240 kW dummy load was connected to the klystron.
In both analog and digital LLRF, the klystron output was controlled by the feedback of the cavity input.  
An output power from the klystron was set to be 20~kW, which is a quarter of the power at the user operation. 
The proportional and integral gains used in the feedback of the digital LLRF were set to be 2.2 and 5.5e4, respectively.

\begin{figure}[hbt]
\centering
\includegraphics[width=60mm]{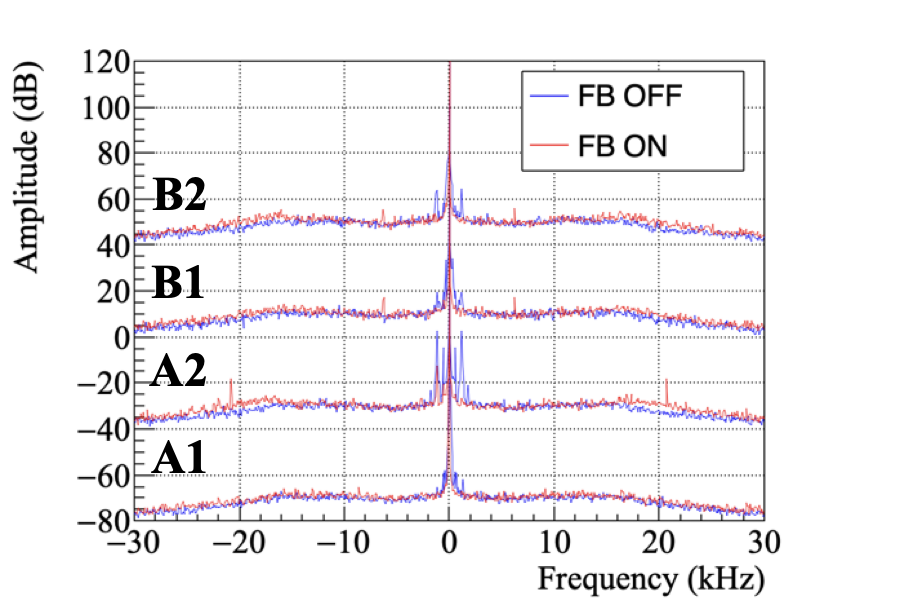}
\caption{The power spectrum of the klystron output controlled by the analog LLRF.}
\label{AnaSP}
\end{figure}

\begin{figure}[hbt]
\centering
\includegraphics[width=60mm]{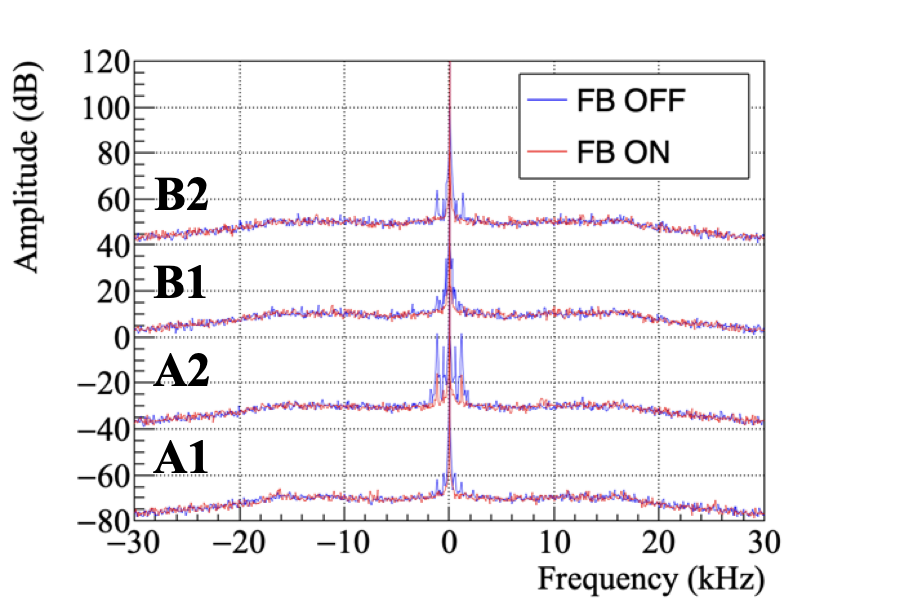}
\caption{The power spectrum of the klystron output controlled by the digital LLRF.}
\label{DigiSP}
\end{figure}

Figures~\ref{AnaSP} and~\ref{DigiSP} show the power spectrum of the klystron output controlled by the analog LLRF and the digital LLRF, respectively.
In both the figures, an offset of 40, 80, and 120~dB was added to the stations A2, B1, and B2, respectively. 
When the power was controlled with the analog feedback, white noise level was slightly higher and some spurious signals were superimposed. 
Then, the feedback with the digital LLRF worked better than that with our analog LLRF.

\subsection{Delay time measurement }
\begin{figure}[hbt]
\centering
\includegraphics[width=75mm]{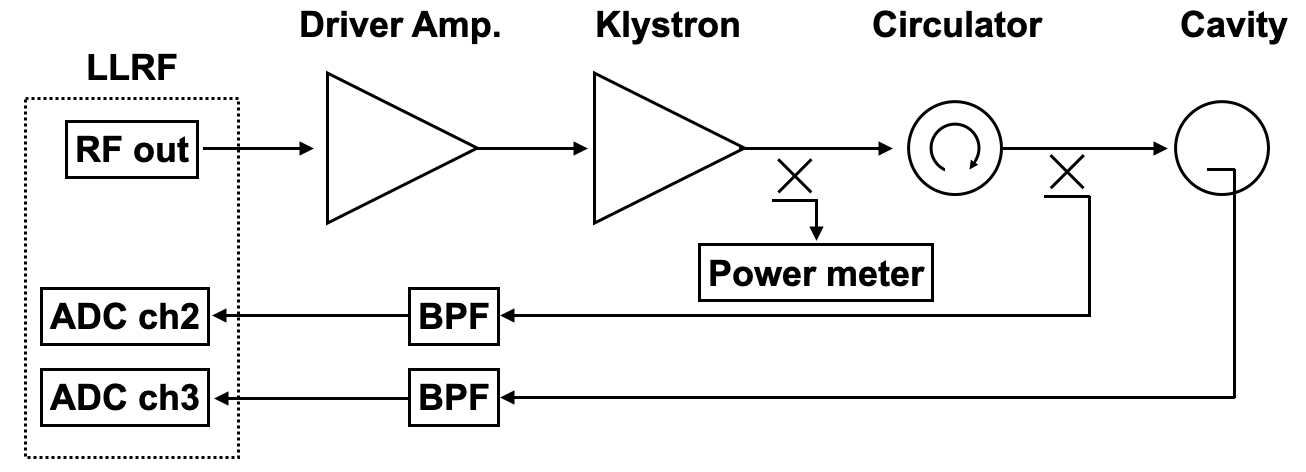}
\caption{Experimental setup of the RF high-power test with the digital LLRF.}
\label{HPwtest}
\end{figure}

We measured the signal delay time of the digital LLRF with the same setup as the user operation, as shown in Fig.~\ref{HPwtest}.
An output RF signal of the LLRF was modulated by a short rectangular pulses, and the signals passing through the RF system were input to the LLRF.
We evaluated the arrival time difference of the waveform, which was recorded by the internal function of the LLRF at each process stage. 
In this measurement, all feedbacks were turned off.

\begin{figure}[hbt]
\centering
\includegraphics[width=80mm]{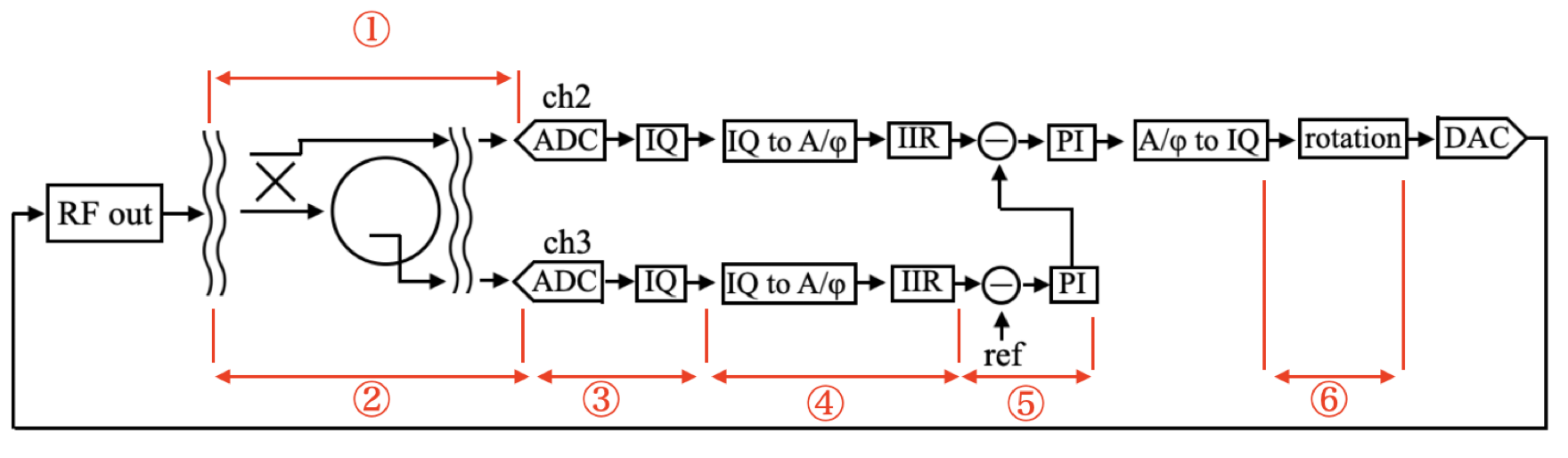}
\caption{Definition of the section ID in the delay time measurement of the digital LLRF.}
\label{MesDef}
\end{figure}

\begin{figure}[hbt]
\centering
\includegraphics[width=60mm]{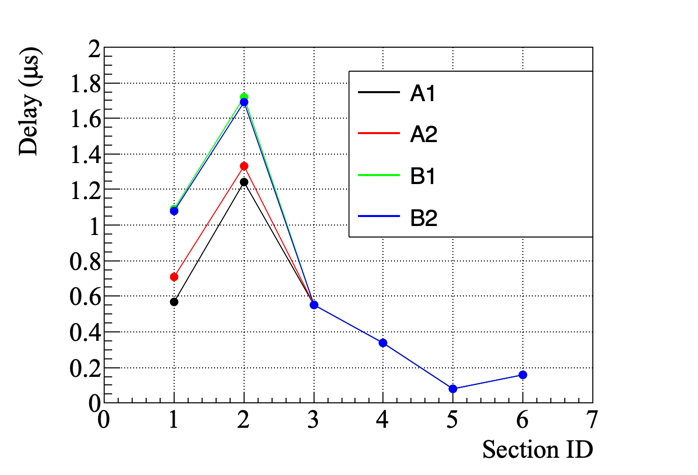}
\caption{The result of the delay measurements in the digital LLRF system.}
\label{Delay}
\end{figure}

Figures~\ref{MesDef} and \ref{Delay} show the definition of the section ID and the delay time at each section, respectively.
From the LLRF output to the LLRF input, the delay time was within 1.2 and 1.7~$\mu$s.
In the LLRF, the delay time of the IQ conversion was the largest and it was caused by the data transfer from the ADC to the FPGA.
The second largest delay was the conversion of the IQ signals into amplitude and phase.
The total delay of the RF system was about 3~$\mu$s, which was acceptable for the use in PF.
If faster feedback is required in the KEK future light source, it is desirable to shorten the delay time.

\subsection{Test of double feedbacks}
We evaluated the stability of the cavity pickup signal with the double feedbacks.
In this test, cavity voltage was set to approximately 410 kV/cavity (power loss of 25 kW) which was close to that under user operations.
The cutoff frequency of the IIR filter just before the cavity input and pickup feedbacks was set to be 10~kHz and 20~Hz, respectively. 
The proportional and integral gains used in the cavity input feedback were fixed to be 2.2 and 2.2e5, respectively.
The proportional and integral gains used in the cavity pickup feedback were swept from 0.5 to 9 and 18 to  5.9e5, respectively.

\begin{figure}[hbt]
\centering
\includegraphics[width=70mm]{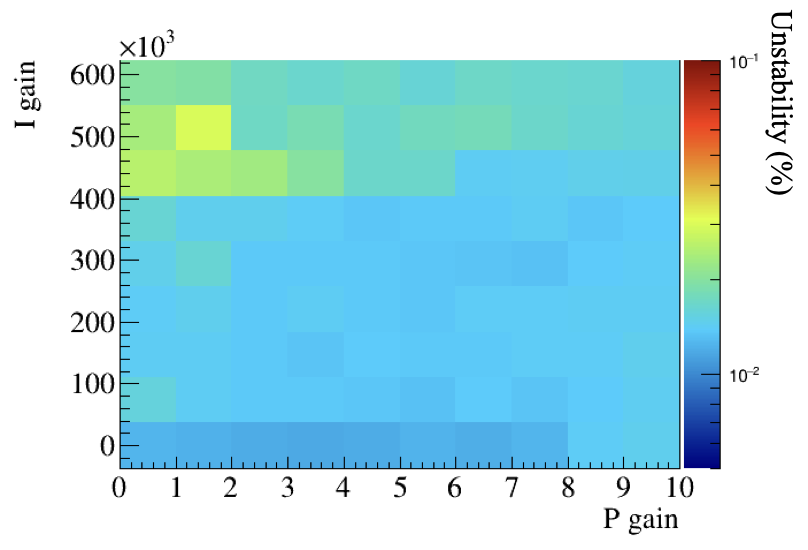}
\caption{Amplitude stability of the cavity pickup signal with respect to the parameters of the proportional and integral gain.}
\label{StaA}
\end{figure}

\begin{figure}[hbt]
\centering
\includegraphics[width=70mm]{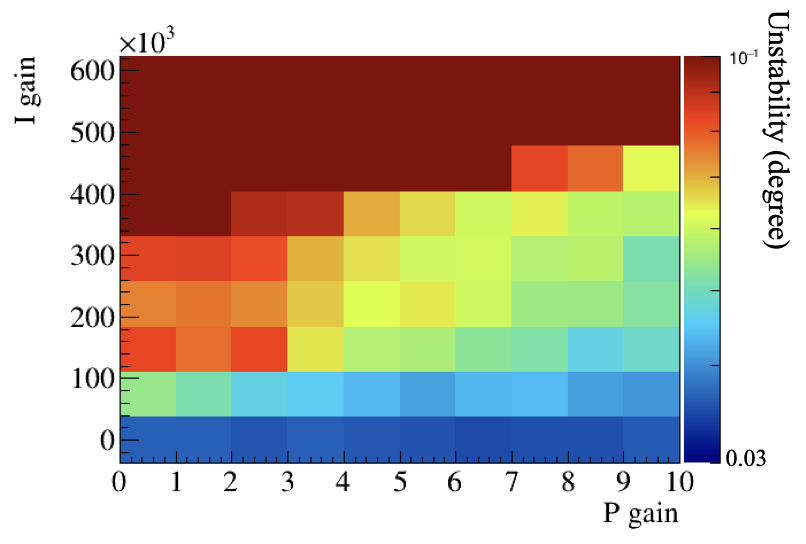}
\caption{Phase stability of the cavity pickup signal with respect to the parameters of the proportional and integral gain.}
\label{StaP}
\end{figure}

Figures~\ref{StaA} and \ref{StaP} show the amplitude and phase stability of the cavity pickup signal with respect to the proportional and the integral gain parameters, respectively.
The horizontal and vertical axis show the set values of the proportional and integral gain, respectively.
The waveform of the cavity pickup signal was recorded at each set of the proportional and integral gains.
Using deviation of the amplitude ($\sigma_a$) and mean of the amplitude ($M_a$) in the single waveform,
the amplitude stability was defined as
\begin{equation}
Stability = \frac{\sigma_{A}}{M_{A}}\times 100~[\%].
\end{equation}
The phase stability was defined as the deviation of the phase in the single waveform.
From Figs.~\ref{StaA} and \ref{StaP}, both amplitude and phase stability were better in the low gain region.
After fine scan of the low gain region, we set the proportional and integral gain to be 2 and 7.3e3, respectively.
With these gain parameters, the amplitude and phase stability were $\pm$0.015~\% and $\pm$0.038 degrees, respectively.
Both the amplitude and phase stability satisfied our goal.

\subsection{Stability tests}
The time dependence of the amplitude and phase stability was measured with the feedback parameters decided in the previous section. 
The amplitude and phase values of the cavity pickup signal were recorded by every 10 minutes.
The reference RF signal was also recorded in order to measure the signal variation inevitably induced in the ADC sampling.

\begin{figure}[hbt]
\centering
\includegraphics[width=70mm]{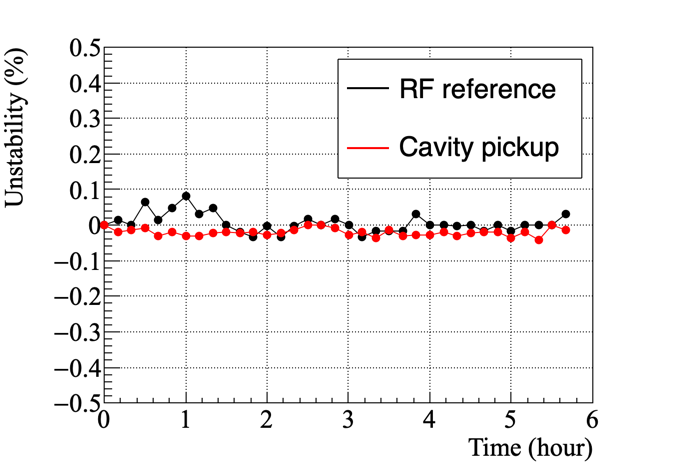}
\caption{Amplitude stability of the cavity pickup signal while it was controlled by the digital LLRF.
 Measured amplitude of reference signal was also shown.}
\label{StaA_t}
\end{figure}

\begin{figure}[hbt]
\centering
\includegraphics[width=70mm]{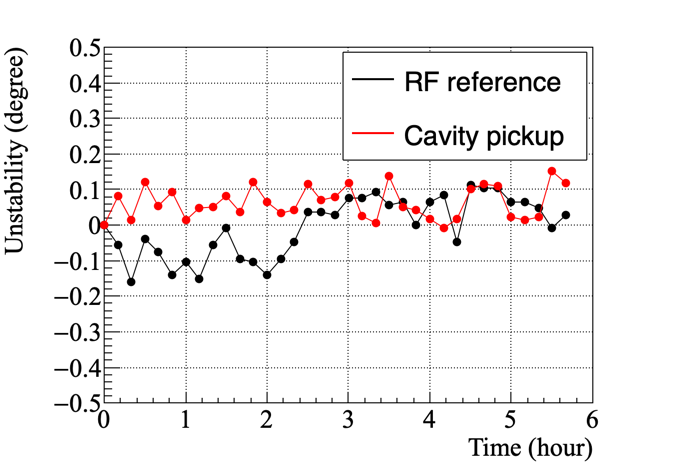}
\caption {Phase stability of the cavity pickup signal while it was controlled by the digital LLRF.
 Measured phase of reference signal was also shown.}
\label{StaP_t}
\end{figure}

Figures~\ref{StaA_t} and \ref{StaP_t} show the amplitude and phase stabilities of the cavity pickup signal, respectively. 
Using the amplitude ($A(t)$) and the phase ($P(t)$) at a time $t$, stability was defined as 
\begin{eqnarray}
Amplitude\ stability = \frac{A(t)-A(0)}{A(0)}\times 100\ [\%],\\
Phase\ stability = P(t)-P(0)\ [^{\circ}].
\end{eqnarray}
From Figs.~\ref{StaA_t} and \ref{StaP_t}, the amplitude and phase stability were found to be $\pm$0.011~\% and $\pm$0.044~degree, respectively. 
Then, we considered the stability observed outside of the LLRF which was degraded by the deviation of the signal sampling at the ADC.
Assuming that the observed deviation of the master RF signal was due to the ADC deviation, 
we calculated the square root of the sum of the cavity pickup and the master RF deviation as the actual stability.
The amplitude and phase stability were calculated to be $\pm$0.03~\% and $\pm$0.09~degree, respectively. 
Though further evaluation of the ADC stability should be performed, 
the amplitude and phase stability were expected to satisfy our goal.

\section{Summary}
We have been developing the digital LLRF based on the MTCA.4 standard for the PF 2.5 GeV ring.
The new LLRF design enable us to detect the transient variation of the cavity voltage, which is the useful technique in a double RF system for KEK future synchrotron light source.  
We demonstrated that the phase variation of the bunches along the bunch train could be measured with our digital LLRF system.
The new LLRF was already installed in the PF ring.
The tentative performance tests of the new LLRF indicated good performance.
In particular, under high power operation, the amplitude and phase stability of the cavity pickup signal were $\pm$0.015~\% and $\pm$0.038 degrees, respectively.
These results satisfied our goal.
Currently, the LLRF system is under commissioning to establish the control parameters and procedures for the user operation.
Then, the commissioning with storage beam will begin in early November.

\section{Acknowledgement}
We thank the following staff of Mitsubishi electric defense and space technologies corporation for their dedication to producing the new LLRF system:  T.~Iwaki, R.~Kitagawa, N.~Terada, T.~Harigae, S.~Yamazaki, and M.~Ryoshi.
We are also grateful to T. Ohshima (SPring-8), F. Tamura (JAEA), T. Kobayashi, Y. Sugiyama, K. Futatsukawa, T. Matsumoto, T. Miura (all of KEK) for useful information and discussions.
\ifboolexpr{bool{jacowbiblatex}}%
	{\printbibliography}%
	{%
	% "biblatex" is not used, go the "manual" way
	
	%\begin{thebibliography}{99}   % Use for  10-99  references
	
} % end \ifboolexpr
%
% for use as JACoW template the inclusion of the ANNEX parts have been commented out
% to generate the complete documentation please remove the "%" of the next two commands
% 
%%%\newpage

%%%\include{annexes-A4}

\end{document}